# On the Tradeoffs of Implementing Randomized Network Coding in Multicast Networks

Yingda Chen and Shalinee Kishore

Electrical and Computer Engineering, Lehigh University

Bethlehem, PA 18015

{yingda, skishore}@lehigh.edu

**Abstract: Randomized network coding (RNC) greatly reduces the complexity of implementing network coding in large-scale, heterogeneous networks. This paper examines two tradeoffs in applying RNC: The first studies how the performance of RNC varies with a node's randomizing capabilities. Specifically, a limited randomized network coding (L-RNC) scheme - in which intermediate nodes perform randomized encoding based on only a limited number of random coefficients - is proposed and its performance bounds are analyzed. Such a L-RNC approach is applicable to networks in which nodes have either limited computation/storage capacity or have ambiguity about downstream edge connectivity (e.g., as in ad hoc sensor networks). A second tradeoff studied here examines the relationship between the reliability and the capacity gains of generalized RNC, i.e., how the outage probability of RNC relates to the transmission rate at the source node. This tradeoff reveals that significant reductions in outage probability are possible when the source transmits deliberately and only slightly below network capacity. This approach provides an effective means to improve the feasibility probability of RNC when the size of the finite field is fixed.**

## I. INTRODUCTION

Network coding enables intermediate nodes in a communication network to combine multiple data packets from incoming links via simple encoding before transmission on outgoing links. It has been shown [1][2] that the *max-flow min-cut* capacity[3][4] of a multicast network can be achieved via linear network coding. Despite this promising result, the construction and maintenance of a rationally-designed network code that achieves *max-flow capacity* can be computationally challenging, especially so for large-scale, dynamic networks. In recognition of such difficulties, Ho *et al.* proposed a *randomized network coding* (RNC) [5][6] scheme that uses random coefficients for encoding. Although suboptimal in general, RNC has two attractive features: First, randomized network codes can be constructed distributively, thereby facilitating the application of network coding in large-scale networks [7]; and second, RNC achieves max-flow capacity with a probability that rapidly approaches 1 as the size of the associated finite field increases [6].

Although randomized encoding considerably reduces the complexity of implementing network coding, efforts continue to further improve its practical applicability and/or to minimize its associated resource costs. For example, Chou *et. al* introduce a practical protocol to implement RNC in [8]. In [9], evolutionary genetic algorithms are used to identify subset of nodes that *have to* perform encoding to achieve desired network capacity (without requiring all nodes be capable of performing encoding). The authors of [10] study a RNC scheme in which nodes choose non-zero random coefficients with a probability $p$. As $p$ decreases, the encoding matrix becomes more "sparse", thereby reducing the encoding/decoding complexity of RNC.

In this paper, we study two important tradeoffs that may arise in applying RNC. The first tradeoff looks at how the performance of RNC varies with a node's randomizing capabilities. To do so, we first propose a *limited randomized network coding* (L-RNC) scheme, in which random combining is performed based on a limited number of random coefficients. In the conventional RNC scheme, when an intermediate node receives $n$ incoming packets (from $n$ incoming edges) and (after encoding) has $m$ outgoing packets to send (on $m$ outgoing edges), the node will generate $n \times m$ random coefficients for each round of encoding. We refer this original RNC scheme [6] with a "*full*" number of encoding coefficients as *exhaustive randomized network coding* (E-RNC). The proposed L-RNC, on the other hand, uses considerably less (i.e., $n$)[1] random coefficients for each round of encoding. Particularly, the $n$ encoding coefficients are *permuted* to encode all $m$ outgoing packets. A direct motivation for this L-RNC scheme is to reduce the computational/storage burden for intermediate nodes; a more subtle motivation is that an intermediate node may not always know the exact value of $m$ in networks with dynamic topologies, e.g., in ad hoc networks where the downstream connectivity can be ambiguous and subject to change. Regardless, we use the L-RNC construct here to quantify how the outage probability of RNC is affected by such limited encoding capabilities of intermediate nodes. We then compare the performance of L-RNC and E-RNC for $n$-dimensional grid networks; both analytical and numerical results suggest that the outage probability of L-RNC is only *slightly* higher than that of E-RNC.

---

[1]Justifications for encoding with $n$ random coefficients, instead of any other number of coefficients between $[1, n \times m)$ will be discussed later in this paper.

Next, we reveal an inherent tradeoff between the capacity gains and the reliability (i.e., the recoverability) of RNC. This tradeoff is achieved when the source intentionally transmits at a rate lower than network-flow capacity so as to achieve lower outage probability at the destination. We note that the notion of transmitting below max-flow capacity was proposed previously in [11]. This earlier study discusses how the size of finite field can impact the feasibility of a *rationally designed* network code in *non-multicast* networks. The results suggested that for certain non-multicast networks, transmitting below capacity may be desirable to avoid an exponentially large finite field. In contrast, our study targets RNC for multicast networks, and the capacity-reliability tradeoff revealed herein represents a fundamentally different line of study from that in [11]. We show through both analytical and simulation results that, *ceteris paribus*, a *small* decrease (from max-flow capacity) in the source rate can provide a *significant* gain in the recoverability of RNC at the destination. This approach provides an effective means to improve the feasibility probability of RNC when the size of the finite field is fixed.

In Section II, we briefly describe the modeling of general randomized network coding and define the limited randomized network coding scheme studied in this paper. The performance of L-RNC is examined in Section III. We then study the tradeoff between capacity and reliability for a general RNC scheme in IV. Our conclusions are presented in Section V.

## II. System Descriptions

### A. Encoding and Decoding of RNC

In this paper, we consider acyclic multicast network represented by a *directed* graph $\mathcal{G}=(\mathcal{N},\mathcal{E})$, such that $\mathcal{N}$ is the set of nodes and $\mathcal{E}$ is the set of the edges. We denote $e_{i,j}\in\mathcal{E}$ as a directed edge from node $i$ to $j$ over which information is sent. Each edge in assumed to be of unit capacity in this study. When confusion can be avoided, a simplified term $e \in \mathcal{E}$ or $\ell \in \mathcal{E}$ is used to denote an arbitrary edge. Nodes $o(e)$ and $d(e)$ are called the origin and destination of an edge $e$, respectively. Edges that share a common origin $i$ form a set $\mathcal{E}_o(i) \subseteq \mathcal{E}$, while $\mathcal{E}_d(i) \subseteq \mathcal{E}$ denotes the set of edges that share a common destination $i$. We study here the *multicast connection* *problem*, in which every receiver is interested in receiving *all* source information. We assume that there is only one source node[2] $\alpha \in \mathcal{N}$ and that it transmits $R$ discrete memoryless information processes from set $\mathcal{S}_R = \{x_1, x_2, \cdots, x_r, \cdots, x_R\}$. The set of receive (sink) nodes is denoted as $\{\beta_1, \beta_2, \cdots, \beta_d, \cdots \beta_D\} = \mathcal{H}_\beta \subset \mathcal{N}$. All other nodes are called intermediate nodes and form the set $\mathcal{I}$; thus, $\mathcal{N} = \{\alpha\} \bigcup \mathcal{I} \bigcup \mathcal{H}_\beta$, where $\{\alpha\}, \mathcal{I}$ and $\mathcal{H}_\beta$ are non-overlapping.

In linear network coding, $y_e$, the information process transmitted on edge $e$, is a linear combination of the information processes carried on the incoming edges of node $o(e)$, i.e.,

$$y_e = \sum_{r \in \mathcal{R}} a_{r,e} x_r + \sum_{\ell \in \mathcal{L}_\ell} f_{\ell,e} y_\ell = \mathbf{a}_e^{\mathrm{T}} \mathbf{x} + \mathbf{f}_e^{\mathrm{T}} \mathbf{y}_\ell, \tag{1}$$

where $\mathcal{R} = \{r \in \mathcal{E} : d(r) = o(e), o(r) = \alpha\}$, $\mathcal{L}_\ell = \{\ell \in \mathcal{E} : d(\ell) = o(e), o(\ell) \in \mathcal{I}\}$, $a_{r,e}$ and $f_{l,e}$ are the local coding coefficients at node $o(e)$. Here we index the source edges with the same subscript $(\cdot)_r$ as the source information processes, since each process is mapped onto one outgoing edge from the source. Under RNC schemes, these coefficients are *randomly* chosen from finite field[3] $\mathbb{F}_{2^u}$. The network code in (1) can be described by the double $(\mathbf{A}, \mathbf{F})$, where $\mathbf{A} = \{a_{r,e}\}$ is a $R \times |\mathcal{E}|$ matrix and $\mathbf{F} = \{f_{\ell,e}\}$ is a $|\mathcal{E}| \times |\mathcal{E}|$ matrix. Since information carried on each edge is a linear combination of source information, we have

$$y_\ell = \mathbf{g}_\ell^{\mathrm{T}} \mathbf{x} \tag{2}$$

for an arbitrary incoming edge $\ell$, where $\mathbf{g}_\ell = [g_\ell^1, g_\ell^2, \cdots g_\ell^R]$ is the global encoding vector (GEV) [8] for edge $\ell$ with respect to (w.r.t.) the source information $\mathbf{x} = [x_1, x_2 \cdots x_R]$. At node $i = d(\ell)$, this information is updated by (1), and we can substitute (2) into (1) to obtain the GEV for an outgoing edge $e$ as $\mathbf{g}_e = \sum_{\ell \in \mathcal{L}_\ell} f_{\ell,e} \mathbf{g}_\ell + \mathbf{a}_e$, i.e., GEV can be updated at each intermediate node.

In RNC scheme, to enable decoding of source information at the receiver, transmitted packets on each edge $e$ will contain the network-encoded information $y_e$ and the GEV $\mathbf{g}_e$. Assuming that a receiver $\beta$ is able to receive $m$ data packets[4] $[\mathbf{g}_1^\beta, y_1], [\mathbf{g}_2^\beta, y_2], \cdots, [\mathbf{g}_m^\beta, y_m]$, a decoding matrix $\mathbf{G}_\beta$ can then be constructed at this receiver, such that $\mathbf{G}_\beta = [\mathbf{g}_1^\beta, \mathbf{g}_2^\beta, \cdots, \mathbf{g}_m^\beta]^{\mathrm{T}}$. The number ($r'_\beta$) of information processes that can be correctly decoded by receiver $\beta$ is equal to $\mathrm{rank}(\mathbf{G}_\beta)$. In the first part of our study, we are interested in the *feasibility probability* of a network code $(\mathbf{A}, \mathbf{F})$ for the multicast connection problem, i.e., the probability that

$$\mathrm{rank}(\mathbf{G}_\beta) = R, \ \forall \beta \in \mathcal{H}_\beta. \tag{3}$$

---

[2]The case of multiple sources can be examined by a straightforward extension that distribute source information to multiple source nodes.

[3]The finite field can be chosen to be of size $p^u$ for any prime number $p$ and integer $u$; we use $\mathbb{F}_{2^u}$ in this study for the convenience of subsequent analysis.

[4]In practice, the number of symbols that can be transmitted in one packet depends on the edge capacity of the specific network. Each data packet may contain a long sequence of symbols from $\mathbb{F}_{2^u}$ that are encoded together, and the encoding vector implies only a relatively small overhead. However, for clarity of analysis, we assume here that each edge is of unit capacity.

This feasibility probability can also be given as $\mathrm{P}_f = 1 - \mathbb{P}_O$, where $\mathbb{P}_O$ is the probability of outage. Outage occurs when *any* receiver fails to recover *any* of the $R$ source processes[5].

When E-RNC is used in multicast networks, each outgoing edge is encoded individually and an intermediate node $i$ generates $|\mathcal{E}_d(i)| \times |\mathcal{E}_o(i)|$ random coefficients from the finite field for *each round* of random combining [5][6]. Generating these random coefficients, however, can be a computational/storage burden to intermediate nodes. The authors in [10] propose a scheme to reduce this burden by *sparse network coding* in which non-zero random coefficients are used at a node only with probability $p$. In our study, we lessen this burden on intermediate nodes by employing limited randomized network coding (L-RNC), which uses only $n = |\mathcal{E}_d(i)|$ random coefficients for each round of encoding. In addition to reducing computational/storage complexity, L-RNC is also motivated by the ambiguity of downstream connectivity (i.e., uncertainty on the value of $|\mathcal{E}_o(i)|$) that may exist in certain (e.g., ad hoc) network topologies. In the original RNC scheme [5][6] and most other studies on RNC, it is implicitly assumed that the quantities $|\mathcal{E}_d(i)|$ and $|\mathcal{E}_o(i)|$ are always known; this is based on the premise that network edges are error-free (or at least that exact information on edge failures, i.e., *erasures*, is known beforehand [12]). However, while the quantity $|\mathcal{E}_d(i)|$ indeed can be readily obtained in most applications, the determination of $|\mathcal{E}_o(i)|$ may be difficult in certain dynamic networks, e.g., due to change in network topologies, instances of downstream edge failures, etc. Even when there is no "physical" change in network topology, if one of the downstream nodes simply refuses to encode the information it received, the *effective* number of downstream edges (i.e., value of $|\mathcal{E}_o(i)|$) will be reduced. Other factors, such as the "reachability" of broadcasting edges in some (wireless) networks can also contribute to the ambiguity of downstream connectivity. Overall, uncertainty about $|\mathcal{E}_o(i)|$ complicates the determination of the (necessary) value of $|\mathcal{E}_d(i)| \times |\mathcal{E}_o(i)|$ needed in E-RNC. Therefore, in such circumstances, the L-RNC approach is more practical. Overall, L-RNC provides a mechanism to quantify the tradeoff between the performance of RNC and a node's randomizing capabilities. It is for these reasons that the L-RNC method is proposed and studied here.

---

[5]We should note that when a network code is said to be "infeasible" (or that an outage has occurred), it indicates that the max-flow of the network is not achieved at all destinations. Some practical "real-time" decoding processes have been discussed (e.g., in [8]) which accumulate "degrees of freedom" in the preparation of decoding to speed up final decoding. Yet decoding of any information is still not guaranteed before $R$ degrees of freedom have been accumulated. Therefore, an outage in a network coding scheme requires retransmission of all $R$ information packets, which could be costly in practice (e.g., for time-sensitive applications). In Section IV, we discuss an alternate approach to reduce such instances of outage when RNC is used.

*B. Limited Randomized Network Coding (L-RNC)*

In the following, we denote the number of random coefficients generated for each round of encoding as $\sigma_i$. In the E-RNC scheme, where downstream connectivity is known and each intermediate node $i$ is capable of generating the "full" number of random coefficients each round, we have $\sigma_i = |\mathcal{E}_d(i)| \times |\mathcal{E}_o(i)|$. As a result, every outgoing link is encoded independently. In L-RNC, we assume that an intermediate node $i$ will use only a "limited" number of random coefficients in each round of encoding. Thus, what is the appropriate set of values of $\sigma_i$ for us to use in studying the L-RNC scheme? To answer this question, we first note that randomized network coding relies on the fact that (despite random combining), the data packets received at the destination become dependent only rarely, i.e., with a probability below some tolerable outage probability. If we select $\sigma_i < |\mathcal{E}_d(i)|$, incoming packets at an intermediate node $i$ will immediately become correlated upon encoding. Although such correlations at an intermediate node do not necessarily imply that the data that eventually propagates to the destination will be dependent (i.e., an outage occurs), the deliberate introduction of correlation contradicts the premise of RNC. For these reasons, proper choice on $\sigma_i$ should satisfy the condition that $\sigma_i \geq |\mathcal{E}_d(i)|$. On the other hand, with the ambiguity on downstream connectivity, intermediate nodes in some networks may not be able to determine exact value of $|\mathcal{E}_o(i)|$, implying that we should choose $\sigma_i$ to be independent of $|\mathcal{E}_o(i)|$. As a result, in the L-RNC scheme discussed in this paper, we assume $\sigma_i = |\mathcal{E}_d(i)|$. With this choice of $\sigma_i$ , we lower bound of the feasibility probability, and (as is shown later in the Proof of Theorem 3.1) we demonstrate that this worst-case L-RNC provides the same theoretical lower bound on the feasibility probability as when $\sigma_i \leq (|\mathcal{E}_d(i)|-1) \times |\mathcal{E}_o(i)|$.

From the discussions above, we now provide a formal definition of the L-RNC:

*Definition 2.1 (Limited Randomized Network Coding):* For an intermediate node $i$ in an acyclic network, let $s_i = |\mathcal{E}_d(i)|$ be the number of incoming information processes received at node $i$

and $t_i = |\mathcal{E}_o(i)|$ be the number of the (distinct) outgoing edges from $i$; we assume $t_i \leq n!$ here. The $s_i$ information processes form an information vector $\mathbf{v}$. In the L-RNC scheme, $s_i$ random coefficients from a finite field $\mathbb{F}_{2^u}$ will be generated to form encoding vector $\mathbf{z}$. Let $\mathbf{z}^{(j)}$ be the $j^{th}$ non-repetitive permutation of $\mathbf{z}$. Node $i$ will transmits the linear encoded information $u^{(j)} = \mathbf{v}^{\mathrm{T}} \mathbf{z}^{(j)}$ (packed with the updated GEV) onto the $j^{th}$ ($j = 1, 2, \cdots t_i$) outgoing edge, thus finishing the L-RNC encoding process at node $i$.

In some special cases, the proposed L-RNC is equivalent to the E-RNC first studied in [5][6]. For example, when $t_i = 1$, we have that only one encoding process is performed in each round and no permutation in $\mathbf{z}$ is required for L-RNC. When $t_i = 1$ for *all* intermediate nodes $i$, the resulting network may look like the one in Fig. 1 (which is similar to the one-dimensional Tandem network studied in [13]). Another (maybe more interesting) scenario when L-RNC and E-RNC are identical is when intermediate nodes "broadcast" [6] the same information to *immediate* downstream nodes. Specifically, when an intermediate node $i$ has multiple outgoing edges, it can choose to transmit the same encoded messages onto different outgoing edges. In such case, the *effective* number of outgoing edges from node $i$ is reduced to $t_i = 1$. In the following, we refer this special case of L-RNC (or E-RNC) as *broadcasting-RNC* (B-RNC), and we will treat B-RNC as a special case of L-RNC in the following discussion.

## III. Performance Analysis for Limited Randomized Network Coding

### A. Performance bounds of L-RNC

Consider a multicast network $\mathcal{G}$ with reliable edges. The network has $D$ receivers and its topology allows a min-cut value of $R$. A RNC $(\mathbf{A}, \mathbf{F})$ is constructed over finite field $\mathbb{F}_{2^u}$. Let $\zeta = \max_i\{|\mathcal{E}_o(i)|\} = \max_i\{t_i\}$ and choose $u$ such that $u > \log_2 D\zeta$. Parameter $\eta$ is defined to be the maximum number of the edges that 1) constitute an edge-disjoint flow solution for the multicast problem; and 2) carry information *generated by random encoding*. We now give a lower bound on the feasibility probability of L-RNC in the following theorem:

---

[6]Note that strictly speaking, the term "broadcast" used here can be more accurately recognized as multicast transmission, since the same information is only transmitted to *immediate* downstream nodes, instead of to all network nodes. However, to avoid confusion with the overall multicast problem, we follow convention and refer to transmitting the same information on all outgoing edges as broadcasting. Similar terms are used in our subsequent discussions on "broadcast sets" and "hyper-edges".

*Theorem 3.1:* When L-RNC is performed in the network topology described above, the lower bound on the network feasibility probability can be given as

$$\mathrm{P}_{LB} = \begin{cases} (1 - \frac{D\zeta}{2^u})^{\frac{\eta}{\zeta}}, & \text{if } t_i | \eta \, , \, \forall i \\ \inf\left\{(1 - \frac{D\mathfrak{z}}{2^u})^{\lceil\frac{\eta}{\mathfrak{z}}\rceil} : \mathfrak{z} \in \mathcal{Z}^+, \mathfrak{z} \leq \zeta\right\}, & \text{if } t_i \nmid \eta \, , \, \exists i \end{cases} \, , \qquad (4)$$

where $\lceil \cdot \rceil$ is the integer ceiling function.

*Proof:* See Appendix A. ∎

We can see that in the special case when $\zeta = 1$, the lower bound on the feasibility possibility achieved by L-RNC is *exactly the same* as that of E-RNC, which was given in [5][6]. In fact, L-RNC and E-RNC achieve the same performance bound only when $\zeta = 1$, which corresponds to a small and special class of networks. When $\zeta > 1$, the feasibility probability of L-RNC will be slightly inferior to that of E-RNC (as shown in the proof of Theorem 3.1 in Appendix A), since the $\mathrm{P}_{LB}$ in (4) achieves its maximum at $\zeta = 1$. In general, at least some intermediate nodes usually have more than 1 distinct outgoing edges. Therefore, the lower bound on the feasibility probability of L-RNC is usually lower than that of E-RNC. However, we note that the feasibility probability of these two schemes both approach 1 exponentially with the length of codeword, and that the gap in the feasibility bounds is quite small, especially when the size of finite field is large enough.

As discussed above, full broadcasting effectively reduces $t_i$ (and therefore $\zeta$) to 1. We now consider the following question: Does L-RNC achieve optimal feasibility when it is used in conjunction with full-broadcasting at intermediate nodes (i.e., when L-RNC is implemented as B-RNC)? Before we answer this question, we should note that grouping different edges into (full or partial) broadcasting "hyperedges" can usually reduce the achievable max-flow in a network [4][14]. For example, in Fig. 2, node $i$ has 3 outgoing edges that are grouped into 2 hyperedges (one of cardinality 2 and the other of cardinality 1). The minimum source-destination cut (i.e., the max-flow capacity of the network) in this case has a value of 2, (as opposed to 3) when the information on each edge is individually encoded and statistically different. Furthermore, if node $i$ adopts the *full* broadcasting strategy, the capacity of this network will be 1. This simple example illustrates how broadcasting can reduce the achievable max-flow of the network; it is therefore generally incorrect to assume that full broadcasting produces optimal feasibility in a network using L-RNC. Nevertheless, for some special networks with high enough degree of "edge redundancy", i.e., where broadcasting does not necessarily reduce network max-flow value, L-RNC can achieve maximum feasibility when used with full broadcasting. In the following, we examine one such class of networks called grid networks.

## B. Performance of L-RNC in Grid Networks

The $2$-dimensional grid network is a rectangular network and was studied as an example in [6]. When $n = 3$, the grid network is a cube-grid shown in Fig. 3. A general $n$-dimensional ($n > 3$) grid network takes the shape of a hyper-cube and can be constructed using $n$ orthogonal axes $V_1, V_2, \cdots, V_n$. Each off-axis intermediate node is connected to $n$ nodes by $n$ incoming edges and $n$ other nodes by $n$ outgoing edges. Further, we assume

- Source node sends data packet $x_i$ on $V_i$ axis, $\forall i = 1, 2, \cdots, n$;
- All on-axis [7] nodes *repetitively relay* incoming information to *all* next hop nodes;
- For L-RNC, an off-axis intermediate node generates $n$ random coefficients for each round of encoding. The $n$ random coefficients are then permuted ($n$ times) to encode the information transmitted on each of the $n$ outgoing edges;
- For B-RNC, an off-axis node generates $n$ random coefficients to linearly encode $n$ incoming information. The encoded information is broadcast onto all $n$ outgoing edges; and finally,
- For E-RNC, $n^2$ random coefficients are generated for each round of encoding and are used to encode information transmitted onto each of the $n$ outgoing edges independently (as per [5][6]).

*Corollary 1:* In the $n$-dimensional grid network described above, a source $\alpha$ is located at origin and a destination $\beta$ is located off-axis at $(\nu_1, \nu_2, \cdots \nu_n)$ (such that $\prod_i^n \nu_i \neq 0$). Let $P_{\text{LB}}^{\text{grid}}$ be the lower bound on the probability that $\beta$ can correctly recover all $n$ source information. When L-RNC is used, $P_{\text{LB}}^{\text{grid}} = \left(1 - \frac{n}{2^u}\right)^{\left(\sum_{i=1}^{n} |\nu_i| - 2\right)}$. When B-RNC is used, $P_{\text{LB}}^{\text{grid}} = \left(1 - \frac{1}{2^u}\right)^{n\left(\sum_{i=1}^{n} |\nu_i| - 2\right)}$.

*Proof:* To prove this corollary, we consider a more general case: $n$ random coefficients are generated for each round of encoding at each intermediate node; each intermediate node groups the $n$ outgoing edges into $J$ hyperedges ($1 \leq J < n$), such that there are $J$ partial broadcasting transmissions (each onto a hyperedge); and the $n$ random coefficients are permuted $J$ times for encoding information transmitted on the $J$ hyperedges.

---

[7]Here, "on-axis" nodes refer to network nodes that have at least one zero coordinate coefficient, i.e., a node $x$ located at $(x_1, x_2, \cdots x_n)$, is an "on-axis" node if its coordinate coefficients satisfy $\prod_{i=1}^{n} x_i = 0$. Similarly, "off-axis" nodes refer to nodes whose coordinate coefficients are all non-zero.

For a $n$-dimensional grid network, there are $n$ *node-disjoint* paths from $\alpha$ to $\beta$; therefore, no matter which broadcasting strategy is chosen at each intermediate node, an edge-disjoint network flow of capacity $n$ can always be found. From the topology of the grid network, we can verify that when $\beta$ is located at $(\nu_1, \nu_2, \cdots \nu_n)$, there can be at most $\sum_{i=1}^{n} |\nu_i| - 2$ intermediate nodes that perform random encoding on each source-destination path. Thus, we have $\eta = n \left( \sum_{i=1}^{n} |\nu_i| - 2 \right)$. Also, we have $D = 1$ and $\zeta = \max\{t_i\} = J$. Applying Theorem 3.1 (or Lemma A.2 in the Appendix) here, and we can obtain $P_{\text{LB}}^{\text{grid}} = \left( 1 - \frac{J}{2^u} \right)^{\lceil \frac{n(\sum_{i=1}^{n} |\nu_i| - 2)}{J} \rceil}$ directly. The special cases of $J = n$ and $J = 1$ correspond to the lower bound on the feasibility probabilities for the L-RNC and B-RNC schemes, respectively. □

Based on Corollary 1, we see that the lower bound on the feasibility probability of B-RNC scheme is identical to that of E-RNC scheme, reported earlier in [5][6] when $n = 2$. That is, for a grid network, E-RNC (which generates $n^2$ random coefficients each round and encodes $n$ outgoing edges differently) is not required to achieve the theoretical maximum lower bound on feasibility for randomized coding. Instead, B-RNC suffices with only $n$ random coefficients and 1 encoding process per round while L-RNC, which encodes information by permuting $n$ random coefficients, performs *slightly* worse than both E-RNC and B-RNC. This *slightly* deteriorated performance of L-RNC is due to information dependencies that may be introduced by the re-use of coefficients. Nevertheless, we should note although B-RNC achieves identical performance to E-RNC (without comprising achievable max-flow capacity) in *this specific network*, broadcasting could reduce network capacity in general topologies and thus may not be desirable in many applications. L-RNC, on the other hand, does not reduce the achievable max-flow of a network and has a feasibility probability bound that is only slightly lower than E-RNC.

The observations above are confirmed using both numerical and simulation results in Fig. 4 for a 3-dimensional grid network. Specifically, we plot the outage probabilities (i.e., $P_{\text{O}}^{\text{grid}} = 1 - P_{\text{feasible}}^{\text{grid}}$) for L-RNC, B-RNC and E-RNC as functions of $u$ (symbol length). The upper bound on such outage probability (derived from the result in Corollary 1) is also shown in Fig. 4 for comparison. The source, located at the origin (see Fig. 3), transmits three independent information processes to a destination located at $(3, 3, 4)$. For E-RNC, each intermediate node encodes each outgoing link individually with $3 \times 3 = 9$ random coefficients. When L-RNC is applied, in each round, an intermediate node generates 3 random coefficients and permutes them to encode each of 3 outgoing edges differently. We can see that in such a 3-dimensional grid network, B-RNC indeed achieves almost identical performance to E-RNC. In comparison, L-RNC exhibits a slightly higher outage probability compared with the other two approaches. However, the performance gap is quite small, and the outage probabilities for all three schemes decrease approximately exponentially with symbol length.

## IV. CAPACITY-RELIABILITY TRADEOFF OF RANDOMIZED NETWORK CODING

When RNC is used, it is possible that the received packets may be unrecoverable at the destination even when all network edges are *error-free*. This is because the random combination performed at different network nodes may introduce dependencies among multiple packets and prevent full decoding at the destination. In this section, we consider how introducing transmission "redundancy" can be potentially useful in *significantly* improving data recoverability in RNC. That is, we show how we may be able to trade some of the capacity gains of randomized network coding for additional reception reliability (i.e., improved feasibility) at the destinations.

Suppose a certain network topology has a network capacity (max-flow) of $R$. If the source sends information at full capacity, all destinations will be able to recover all $R$ information processes with a feasibility probability lower bounded[8] by (4). When a destination $\beta$ fails to accumulate enough degrees to freedom, i.e., when $r'_\beta = \text{rank}\{\mathbf{G}_\beta\} < R$ and an outage occurs, it is generally impossible for it to decode *any* of the received information processes. As a result, the source will have to retransmit *all* $R$ information packets. To reduce this cost of an outage, we pursue here the possibility of improving information recoverability at the destination by slightly reducing the source transmission rate. To show that such a tradeoff between capacity and information recoverability can be made in RNC schemes, we first provide a measure of reliability gain when the source transmits below network capacity:

*Definition 4.1:* (*Reliability Gain for RNC*) When information can be reliably transmitted over network edges and RNC is applied, the reliability gain of RNC w.r.t. outage probability is defined as

$$\mathcal{D}_\mu^v = \lim_{u \to \infty} \frac{\log_2 \mathbb{P}_v}{\log_2 \mathbb{P}_\mu},$$ (5)

---

[8]Or equivalently, by a probability of outage that is upper bounded by one minus the probability in (4).

where $\mathbb{P}_\mu$ is the upper bound on outage probability when the source transmits at capacity $\mu$ and $\mathbb{P}_\upsilon$ is the upper bound on outage probability when the source transmits at a reduced capacity $\upsilon$, such that[9] $1 < \upsilon < \mu \leq R$.

The concept of reliability gain here resembles (in a way) the concept of diversity [15] widely used in wireless communications; both describe criteria that measure the "relative" possibility of recovering source information at the destination. The difference is that diversity is used to combat link failure due to wireless fading; RNC reliability gain is used here to reduce outage that arises from possible data dependency due to random combining, which can happen even when all network edges are reliable[10]. We now study the reliability gain for a multicast network in the following theorem:

*Theorem 4.1:* Consider a multicast network $\mathcal{G}_R$ with reliable edges and a min-cut value of $R$. The source transmits $R$ dependent information processes *expanded by* $Q$ ($1 < Q < R$) independent information processes. The formation of the dependent information processes are performed such that any $Q$ (out of $R$) information processes transmitted by the source must be independent of each other. When intermediate nodes perform E-RNC, a reliability gain of $\mathcal{D}_R^Q = \frac{\eta_Q}{\eta_R} \cdot \binom{R}{Q}$ can be achieved. With L-RNC and $t_i | \{\eta_R, \eta_Q\}$, a reliability gain of $\mathcal{D}_R^Q = \frac{\eta_Q \zeta_R}{\eta_R \zeta_Q} \cdot \binom{R}{Q}$ can be achieved, where $\eta_R, \zeta_R$ are the parameters defined in Theorem 3.1 and $\eta_Q, \zeta_Q$ are equivalent parameters for a reduced network $\mathcal{G}_Q$ obtained by deleting certain edges from $\mathcal{G}_R$, e.g., by deleting $R - Q$ edges in a max-flow cut, till a max-flow of $Q$ is obtained.

*Proof:* See Appendix B. ∎

Theorem 4.1 provides a capacity-reliability tradeoff when a network of $R$ disjoint source-destination paths is used to transmit $Q < R$ independent information processes. Further, it shows that the recoverability of randomly encoded information can be *significantly* improved by only slightly decreasing the transmit rate. Although the network outage probability could be reduced by increasing size of finite field $\mathbb{F}_{2^u}$ as well, the arithmetic operations in a larger finite field are more complicated and may not always be desirable. In contrast, Theorem 4.1 provides an alternate way of improving network reliability with a small cost in network capacity. In fact, when either E-RNC or L-RNC is used, a reliability gain close to $\binom{R}{Q}$ is usually achievable (depending on the specific network topology). We show this next for $n$-dimensional grid network.

---

[9]When $\upsilon = 1$, each intermediate node can perform RNC without introducing local dependency; thus $\mathbb{P}_\upsilon = 0$.

[10]That being said, it is possible to explore this tradeoff between capacity and reliability as a means to counter the impact of fading (i.e., unreliable edges) on the performance of RNC in *wireless* networks. We show in [16] that by scaling back the source transmission rate, higher feasibility can be achieved in a wireless network with fading links.

*Corollary 2:* When L-RNC is applied to the $n$-dimensional grid-network with full broadcasting (i.e., B-RNC) and assuming the source transmits $n$ information processes that are linear combinations of $m$ $(1 < m < n)$ *independent* processes (i.e., $R = n$ and $Q = m$), a reliability gain of $\mathcal{D}_n^m = \binom{n-1}{m-1}$ can be achieved.

*Proof:* For the network topology studied here, it is straightforward to show that $\zeta_R = \zeta_Q = 1$. Also, from the definition of $\eta$ in Theorem 3.1, we have $\eta_R = n(\sum_{i=1}^n \nu_i - 2)$ and $\eta_Q = m(\sum_{i=1}^n \nu_i - 2)$. Applying Theorem 4.1 yields

$$\mathcal{D}_n^m = \frac{m}{n} \cdot \binom{n}{m} = \frac{m}{n} \cdot \frac{n!}{m!(n-m)!} = \frac{(n-1)!}{(m-1)!(n-m)!} = \binom{n-1}{m-1}. \tag{6}$$

For L-RNC and E-RNC schemes, we can derive similar reliability gain from Theorem 4.1. □

As an example, we illustrate the tradeoff between reliability and capacity for a 3-dimensional grid network in Fig. 5. Both theoretical upperbounds as well as simulation results are provided. In our simulation, we assume the destination is located at $(3, 3, 4)$. When full capacity is achieved, 3 independent information processes are transmitted by the source; when the network is operating at reduced capacity (of 2), the source transmits 2 independent information processes and 1 dependent process. We can see that when the source transmits at reduced capacity, the outage probability exhibits a reliability gain of $\binom{3-1}{2-1} = 2$ when the size of the finite field is large. Note that the RNC reliability gain in Fig. 5 is seen as the increase (i.e., doubling) of the slope in the outage probability for large finite fields. We can also see from Fig. 5 that the actual outage probabilities for the 3-dimensional grid network are much lower than the corresponding theoretical upper bounds, both for full and reduced capacity transmission. This is to be expected since the theoretical results derived in Theorems 3.1 and 4.1 (and shown in Fig. 5 ) are the upper bounds on outage probability (i.e., the "worst-case" probability) for any *arbitrary* multicast network. When we choose a specific network topology (especially, a regular, small-scale configuration such as in our simulations), we can expect the actual outage probability to be much lower than the upper bound. However, we see at the same time that the simulation results are asymptotically consistent with the theoretical results: Both show an exponential decay in network outage probability with respect to the length of the transmitted symbol.

We see here that a considerable improvement in feasibility probability can be achieved by trading off a small fraction (especially for networks with large max-flow) of achievable capacity, which may be very desirable in error/delay sensitive applications. Nevertheless, we should also note that network coding was proposed to achieve the max-flow network capacity which may be unachievable otherwise in a general network. A severe reduction in capacity (for improved feasibility) may therefore be unadvisable, especially in networks with reliable edges; instead, a careful decision should be made based on the specific application.

## V. Concluding Remarks and Future Works

In this paper, we study a randomized network coding scheme based on a limited number of random coefficients. The tradeoff between the reduction of the number of random coefficients and feasibility bounds of such L-RNC scheme is studied. Then, a distinct capacity-reliability tradeoff for general RNC schemes is revealed and quantified. This tradeoff provides an alternate approach to improving the probability of recovering randomly-encoded information at the destination without using an excessively large finite field. In both cases, we illustrate the application of the two tradeoffs in the context of grid networks.

While the two tradeoffs stem from different motivations in regards to the implementation of randomized network codes, their outcomes may converge in some special scenarios. For example, in a network where the source node is unaware of max-flow capacity, it may adopt L-RNC and transmit at a rate $Q$ that turns out to be below the max-flow rate $R$. If we make a cut much further down the network, i.e., after multiple tiers of relays, the edges in such cut *may* carry information that satisfy the conditions in Theorem 4.1. In such special cases, the outcome may coincidentally mimic the capacity-reliability tradeoff, i.e., although the source conservatively transmits below max-flow due to downstream edge uncertainty (a motivation for L-RNC), the process of randomized network coding at intermediate nodes *may* produce reliability gains. Although such outcomes are not applicable to arbitrary multicast networks, and in many cases, conservative transmission from the source cannot guarantee the full capacity-reliability tradeoff in Theorem 4.1, it is worth pointing out that the two tradeoffs studied in this work do share a natural connection.

One final remarks we would like to make is that although tradeoffs discussed in this paper are derived based on the model of wireline networks, some of them may carry important implications in wireless settings as well. For example, the tradeoff between capacity and reliability can be explored to reduce outage due to fading in wireless channels. Additionally, the broadcast-RNC approach is a natural fit in the inherently-broadcasting wireless media. We continue this line of work in [16] and explore the application of these tradeoffs in wireless networks with fading and broadcasting edges.

## Appendix

### A. Proof of Theorem 3.1

We first present a lemma[11] that studies the upper bound of the probability that a polynomial over a finite field equals $0$.

*Lemma A.1:* For a polynomial $\mathcal{P}(z_1, z_2, \cdots)$ over finite field $\mathbb{F}_{2^u}$ of total degree $M$, if the maximum exponent in *any* variable $z_i$ is $m$ ($m < M, m < 2^u$), the probability of $\mathcal{P}$ equals zero is upper bounded by $1 - (1 - \frac{m}{2^u})^{\lceil \frac{M}{m} \rceil}$.

*Proof:* Let $m_j$ be the largest exponent of $z_j$ in $\mathcal{P}(z_1, z_2, \cdots)$. The polynomial $\mathcal{P}$ can then be written as $\mathcal{P} = z_1^{m_1} \mathcal{P}_1^c + \mathcal{R}_1$, where $\mathcal{P}_1^c$ is a polynomial that does not contain terms of $z_1$ and has degree no higher than $M - m_1$ and $\mathcal{R}_1$ is the "residue" polynomial of $\mathcal{P}$ w.r.t. $z_1^{m_1}$. From the rules of deferred decision and Schwartz-Zippel theorem [17], we then have $\Pr(\mathcal{P} = 0) \leq \Pr(\mathcal{P}_1 = 0) \left(1 - \frac{m_1}{2^u}\right) + \frac{m_1}{2^u}$. Applying Schwartz-Zippel theorem [17] recursively, we obtain

$$\Pr(\mathcal{P}(z_1, z_2, \cdots) = 0) \leq \frac{\sum_{i=1}^{M} m_i}{2^u} - \frac{\sum_{i \neq j} m_i m_j}{2^{2u}} + \cdots + (-1)^{M-1} \frac{\prod_{i=1}^{M} m_i}{2^{M\mu}}. \tag{7}$$

1) When $\mod (M, m) = 0$. The upper bound on $\Pr(\mathcal{P}(z_1, z_2, \cdots) = 0)$ can be found by solving the integer optimization problem:

$$\text{Maximize} \quad P_{UB} = \sum_{k=1}^{M} \frac{(-1)^{k-1}}{2^{uk}} \left( \sum_{i_p \neq i_q, \forall \{p,q\} \in [1,k]} m_{i_1} m_{i_2} \cdots m_{i_k} \right) \tag{8}$$

$$\text{Subject to} \quad \sum_{j=1}^{M} m_j \leq M$$

$$0 \leq m_j \leq m, \forall j \text{ and } m_j \in \mathbb{Z}$$

Problem (8) can be solved by induction. We assume the optimal solution to (8) is $\mathbf{m}^* = \{m_1^*, m_2^*, \cdots m_M^*\}$. First, we inspect the case when the first constraint $\sum_{j=1}^{M} m_j \leq M$ is *not* binding (i.e., the optimal solution is obtained when $\sum_{j=1}^{M} m_j^* < M$). Then we can always find a feasible solution $\mathbf{m} = \mathbf{m}^* + \epsilon$, where $\epsilon = \{\epsilon_1, \epsilon_2, \cdots \epsilon_M\}$ and $\epsilon_i > 0$, such that $P_{UB}(\mathbf{m} = \mathbf{m}^* + \epsilon) - P_{UB}(\mathbf{m} = \mathbf{m}^*) > 0$. This contradicts the optimality of $\mathbf{m}^*$. Therefore, the first constraint must be binding and we have $\sum_{j=1}^{M} m_j = M$. Similarly, we can show that *both ends* of the second constraint must be binding, i.e. $m_i = 0$ or $m_i = m, \forall i$. Therefore, the optimal (maximal) value of $P_{UB}$ is achieved when exactly $\frac{M}{m}$ of $m_i$ take on the value of $m$ and the remaining $m_i$ take on the value of 0. It is easy to see that this optimal solution yields maximal value of $P_{UB} = 1 - (1 - \frac{m}{2^u})^{\frac{M}{m}}$.

---

[11]The proof of a special case of Lemma A.1 can be found in [6]. We use here a similar technique as the one adopted by authors of [6]. The sketch of the major supporting argument is included for completeness of the proof.

2) When $\mod(M, m) \neq 0$. Consider a polynomial $\mathcal{P}'(z_1, z_2, \cdots)$ in which $z_i$ is of the same maximum degree $m$ and $\mathcal{P}'$ is of maximum degree $M' = m \cdot \lceil \frac{M}{m} \rceil > M$. Again by Schwartz-Zippel Theorem [17], we can show that $\Pr\{\mathcal{P}(z_1, z_2, \cdots) = 0\} < \Pr\{\mathcal{P}'(z_1, z_2, \cdots) = 0\} = 1 - (1 - \frac{m}{2^u})^{\lceil \frac{M}{m} \rceil}$. Therefore, the probability that $\mathcal{P}$ equals zero is upper bounded by $1 - (1 - \frac{m}{2^u})^{\lceil \frac{M}{m} \rceil}, \forall m < M$.

Combination of both cases concludes the proof of this lemma. $\qquad \square$

Now we consider a lemma that, when L-RNC is applied, represents a special (and simplified) case of Theorem 3.1.

*Lemma A.2:* For a multicast problem considered in Theorem 3.1, if $t_i = K$, $\forall i \in \mathcal{I}$, the probability that $(\mathbf{A}, \mathbf{F})$ is feasible is lower bounded by $(1 - \frac{DK}{2^u})^{\lceil \frac{n}{K} \rceil}$.

*Proof:* Assume that each intermediate node $i$ in the network generates $s_i = |\mathcal{E}_d(i)|$ random coding coefficients for each round of relaying. The network has a max-flow of $R$ (if there are broadcasting edges, this max-flow is obtained w.r.t. hyperedge). By max-flow min-cut theorem, we can show that [4] the flow solution for any destination $\beta$ in such network contains $R$ *edge-disjoint* paths $\mathcal{E}_1, \mathcal{E}_2, \cdots, \mathcal{E}_r, \cdots, \mathcal{E}_R$, each connecting a different source process to $\beta$. For a randomized network code $(\mathbf{A}, \mathbf{F})$ to be feasible (i.e., achieve the network capacity $R$), (3) must hold. This means that for the $w_\beta$ information packets received by $\beta$, we must have 1) $w_\beta \geq R, \forall \beta$; and 2) the $R \times w_\beta$ matrix $\mathbf{G}_\beta$ can be reduced to a $R \times R$ matrix $\mathbf{G}'_\beta$ such that $\det(\mathbf{G}'_\beta) \neq 0$, $\forall \beta \in \mathcal{H}_\beta$.

Let $\Theta'_\beta = \det(\mathbf{G}'_\beta)$ be the determinant polynomial for the decoding matrix at destination $\beta$. Using similar arguments as those in [18], we can show that $\Theta'_\beta \neq 0$ if and only if $\Theta_\beta = \sum_r |\mathbf{A}_r| \cdot \prod_{r=1}^{R} g(\mathcal{E}_r^\beta) \neq 0$, where $\mathcal{E}_r^\beta$ is an edge-disjoint path from the source to destination $\beta$; $\mathbf{A}_r$ is a sub-matrix of $\mathbf{A}$ that maps $\mathbf{x}$ onto each path $\mathcal{E}_r^\beta$; and $g(\mathcal{E}_r^\beta)$ is the product of random coefficients along path $\mathcal{E}_r^\beta$. From the definition of $\eta$, we see that $\Theta_\beta$ is a random polynomial of maximum degree $\eta$. With L-RNC, each intermediate node $i$ will encode the $|\mathcal{E}_d(i)|$ incoming information packets with $|\mathcal{E}_d(i)|$ random coefficients and transmit onto $t_i$ edges (or hyperedges)[12]. Although at most one edge from each broadcasting set will appear in a flow solution, the same random coding coefficients $\{a_{r,e}, f_{\ell,e}\}$ (generated at node $i$)can appear at most $\min(R, t_i)$ times in the $R$ edge-disjoint paths. In this lemma we assume $t_i = K \leq R, \forall i$[13], therefore the maximum number of times that $\{a_{r,e}$ or $f_{\ell,e}\}$ can appear in $\Theta_\beta$ is[14] $\zeta' = \min(K, R) = K$. Thus, $\Theta_\beta$ is a polynomial of at most degree $\eta$, in which the maximum exponent of a random variable is $K$.

With multiple receivers, the condition under which all receivers can decode all source processes is $\Theta_{\mathcal{H}_\beta} = \mathrm{C}\prod_{\beta \in \mathcal{H}_\beta} \Theta_\beta \neq 0$, where $\mathrm{C}$ is a constant number and $\Theta_{\mathcal{H}_\beta}$ is the *total determinant polynomial*. From the properties of $\Theta_\beta$, we see that $\Theta_{\mathcal{H}_\beta}$ is a polynomial over $\mathbb{F}_{2^u}$ of maximum degree $D\eta$ in random variables $\{a_r, f_{\ell,e}\}$ and the largest exponent of the random variables is $DK$. Using the result of Lemma A.1, we can derive the lower bound on the feasibility probability, i.e., the lower bound on $\Pr(\Theta_{\mathcal{H}_\beta} \neq 0)$ to be

$$
\begin{aligned}
\mathrm{P}_{LB} &= \inf\left\{\mathrm{P}\left(\Theta_{\mathcal{H}_\beta} \neq 0\right)\right\} \\
&\geq 1 - \left(1 - \left(1 - \frac{DK}{2^u}\right)^{\lceil \frac{D\eta}{DK} \rceil}\right) \\
&= \left(1 - \frac{DK}{2^u}\right)^{\lceil \frac{\eta}{K} \rceil}.
\end{aligned}
\tag{9}
$$

This gives the feasibility probability for a multicast network when L-RNC is used. $\qquad\square$

With Lemma A.2, we can now complete the proof of Theorem 3.1. Recall that $t_i = |\mathcal{E}_o(i)|$ denotes the number of outgoing links from node $i$. Using the result in Lemma A.2, we can immediately show that

---

[12]We implicitly assume that $t_i \leq s_i!$ (factorial of $s_i$), such that the encoding coefficients associated with different broadcast sets are obtained using *non-repetitive* permutations of the $t_i$ random coefficients, as per the definition Definition 2.1.

[13]The relaxation of this assumption will give similar results; but will involve more special cases to consider.

[14]Moreover, in the worst-case scenario (although with low probability), the $K$ non-repetitive permutations of $\sigma = |\mathcal{E}_o(i)| \times |\mathcal{E}_d(i)| - K$ can still have the recurrence of a specific random variable (in the final decoding polynomial) a maximum of $\min(K, R) = K$ times. Also, since $|\mathcal{E}_o(i)| \geq J_i$, we can conclude that the final decoding polynomial remains a polynomial of maximum order $K$ w.r.t. a specific random variable for any $\sigma_i \leq |\mathcal{E}_o(i)| \times (|\mathcal{E}_d(i)| - 1)$.

**Case A**: If $t_i | \eta$ , $\forall i$, let $\frac{2^u}{D} > \lambda \geq 1$, we can define a function $F(\cdot)$ to be $F(\lambda, \eta, D) \equiv (1 - \frac{D\lambda}{2^u})^{\frac{\eta}{\lambda}}$. Taking the derivative w.r.t. $\lambda$, we would have $\frac{\partial F}{\partial \lambda} = \frac{\eta}{\lambda} \cdot F(\lambda, \eta, D) \cdot \mathcal{Y}(\lambda, D)$, where

$$\mathcal{Y}(\lambda, D) = \frac{\ln\left(\frac{2^u}{2^u - D\lambda}\right)}{\lambda} - \frac{D}{2^u - D\lambda} = \mathcal{Y}_1(\lambda, D) - \mathcal{Y}_2(\lambda, D).$$

It is straightforward to show that $\frac{\partial \mathcal{Y}_1}{\partial \lambda} < 0$ and $\frac{\partial \mathcal{Y}_2}{\partial \lambda} > 0$; therefore, we have

$$\max\left\{\mathcal{Y}(\lambda, D) : \lambda \in [1, \frac{2^u}{D})\right\} = \mathcal{Y}(\lambda, D)|_{\lambda = 1} = \ln(1 + \chi) - \chi,$$

where $\chi = \frac{D}{2^u - D} > 0$. By Taylor expansion, it can be shown that $\ln(1 + \chi) < \chi$, $\forall \chi > 0$. Thereby $\max\{\mathcal{Y}(\lambda, D)\} > 0$ for any value of $\lambda$ of interest here. Thus, we have $\frac{\partial F}{\partial \lambda} < 0$, for any value of $\{\lambda, \eta, D\}$ considered in this problem; and that $F(\lambda, \eta, D)$ is a *monotonically decreasing* function of $\lambda$. This conclusion (derived for a continuous $\lambda$) also holds for integer problem in which $\lambda$ is replaced by $\zeta$. Therefore, the feasibility probability in this case is lower bounded by $(1 - \frac{D\zeta}{2^u})^{\frac{\eta}{\zeta}}$.

**Case B**: If $t_i \nmid \eta$ , $\exists i$, we again define $F_2(\cdot)$ to be $F_2(\lambda, \eta, D) = (1 - \frac{D\lambda}{2^u})^{\lceil \frac{\eta}{\lambda} \rceil}$. Unfortunately, $F_2(\cdot)$ is not strictly monotone [15] w.r.t. $\lambda$, due to the existence of the integer ceiling function. Thus, we can only lower bound the feasibility probability by $\inf\{(1 - \frac{D\mathfrak{m}}{2^u})^{\lceil \frac{\eta}{\mathfrak{m}} \rceil} : \mathfrak{m} \in \mathcal{Z}^+, \mathfrak{m} \leq \zeta\}$.

The combination of both cases concludes the proof. ∎

### B. Proof of Theorem 4.1

In the following Lemma, we first consider the how outage probability can be reduced when a network with one single receiver is operating at reduced capacity $Q$.

*Lemma A.3:* Consider a single-receiver network with reliable edges and min-cut value of $R$. Assume that the source transmits at reduced network capacity $Q$ as stated in Theorem 4.1. When E-RNC is used, the probability that the receiver fails to recover *all* $Q$ independent source processes is $\mathbb{P}_Q = \left(1 - (1 - \frac{1}{2^u})^{\eta_Q}\right)^\Gamma$; when L-RNC is used and $t_i | \{\eta_R, \eta_Q\}$, $\mathbb{P}_Q = \left(1 - (1 - \frac{\zeta_Q}{2^u})^{\frac{\eta_Q}{\zeta_Q}}\right)^\Gamma$, where $\Gamma = \binom{R}{Q}$.

*Proof:* First we consider the scenario where L-RNC is applied. By max-flow min-cut theorem, any feasible flow-solution for a network of capacity $R$ comprises $R$ *disjoint* source-receiver paths and therefore $|\mathcal{E}_d(\beta)| \geq R$ at the receiver $\beta$. Now consider the case when $|\mathcal{E}_d(\beta)|{=}R$ (if $|\mathcal{E}_d(\beta)| > R$, we can always reduce the network to an equivalent network such that $|\mathcal{E}_d(\beta)| = R$). In this case, the decoding matrix can be written as $\mathbf{G}_\beta = [\mathbf{g}_1, \mathbf{g}_2, \cdots, \mathbf{g}_R]$. If we arbitrarily delete $R - Q$ edges from $\mathcal{E}_d(\beta)$, the network capacity is immediately reduced to $Q$, and (by definition) there must exist $Q$ disjoint source-receiver paths in this reduced network. The receiver now has a new decoding matrix $\mathbf{G}'_\beta = [\mathbf{g}'_1, \mathbf{g}'_2, \cdots, \mathbf{g}'_Q]$ and we have from Theorem 3.1 that the upper bound on the outage probability for the reduced network is $p_Q = \mathrm{P}\left\{\det(\mathbf{G}'_\beta) = 0\right\} \leq 1 - (1 - \frac{\zeta_Q}{2^u})^{\frac{\eta_Q}{\zeta_Q}}$. Since *every* non-repetitive choice of deleting $R - Q$ of the $R$ edges from $\mathcal{E}_d(\beta)$ will result in a *distinct* flow-solution of capacity $Q$ and since all information transmitted across the network is expanded by $Q$ independent source information, the outage probability that $\beta$ cannot recover *all* $Q$ information processes is simply $\mathbb{P}_Q = \left(1 - (1 - \frac{\zeta_Q}{2^u})^{\frac{\eta_Q}{\zeta_Q}}\right)^{\Gamma}$, where $\Gamma = \binom{R}{Q}$. For the scenario when E-RNC is applied, we can similarly prove that $\mathbb{P}_Q = \left(1 - (1 - \frac{1}{2^u})^{\eta_Q}\right)^{\Gamma}$. □

---

[15]However, following arguments similar to Case A, we can show that the value of $F_2(\cdot)$ *tends to* decrease as $\lambda$ increases. This can serve as a general guideline when considering design tradeoffs in practical networks.

We now consider the reliability gain for multicast networks with multiple receivers $\beta \in \mathcal{H}_\beta$. If the source transmits at a reduced capacity of $Q$, an outage occurs when any one of the $D$ receivers fails to recover all $Q$ independent source processes. The RNC reliability gain in this case can be computed from (5) and using Lemma A.3 as

$$
\begin{aligned}
\mathcal{D}_R^Q &= \lim_{u \to \infty} \frac{\log_2\left(\sum_{\beta \in \mathcal{H}_\beta} \mathbb{P}_Q^\beta\right)}{\log_2 \mathbb{P}_R} \\
&= \lim_{u \to \infty} \Gamma \cdot \frac{\log_2 D + \log_2\left(1 - (1 - \frac{\zeta_Q}{2^u})^{\frac{\eta_Q}{\zeta_Q}}\right)}{\log_2\left(1 - (1 - \frac{D\zeta_R}{2^u})^{\frac{\eta_R}{\zeta_R}}\right)} \\
&\simeq \Gamma \cdot \lim_{u \to \infty} \frac{\log_2 D + \frac{\eta_Q}{\zeta_Q}\left(\log_2(\zeta_Q) - u\right)}{\frac{\eta_R}{\zeta_R}\left(\log_2(\zeta_R) + \log_2 D - u\right)} = \frac{\eta_Q \zeta_R}{\eta_R \zeta_Q} \cdot \Gamma.
\end{aligned}
\tag{10}
$$

When E-RNC is used, we can similarly show that a reliability gain of $\mathcal{D}_R^Q = \frac{\eta_Q}{\eta_R} \cdot \binom{R}{Q}$ can be achieved. ∎

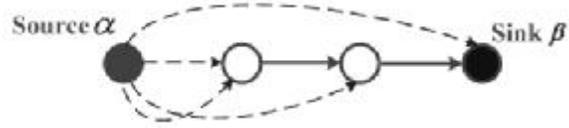

Fig. 1. A tandem network in which only the source has multiple outgoing edges; here, the E-RNC and L-RNC schemes are identical.

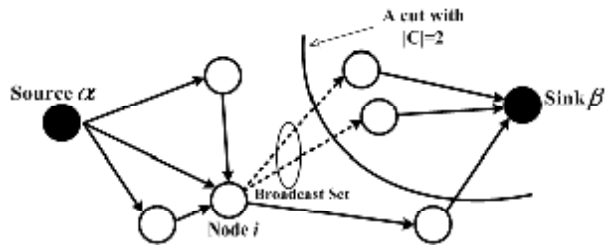

Fig. 2. Example network with broadcasting intermediate nodes.

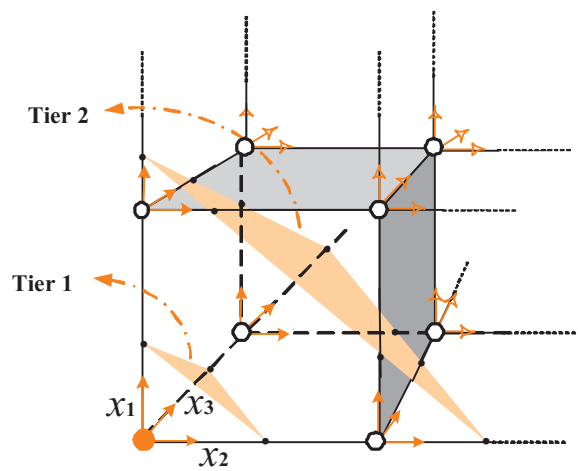

Fig. 3. Three-dimensional grid network.

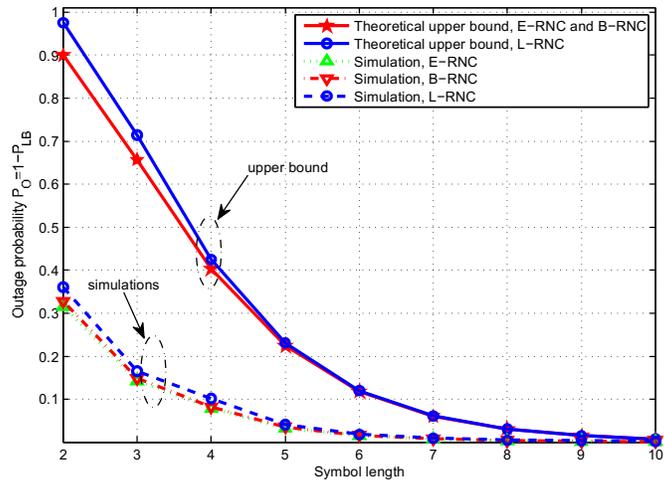

Fig. 4. Performance comparison between randomized network coding with and without full broadcasting.

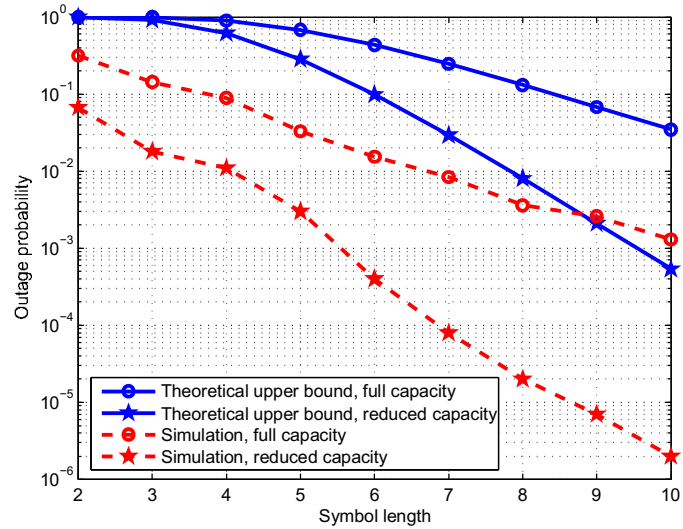

Fig. 5. Reliability and capacity tradeoff for a 3-dimensional grid network.